\documentclass[review,number,sort&compress]{elsarticle}

\usepackage[utf8]{inputenc} 
\usepackage[english]{babel}
\usepackage{graphicx}
\usepackage{hyperref}
\usepackage{subfig}

\graphicspath{{./figures/}}

\begin{document}

\begin{frontmatter}
  
\title{Deep Diffused Avalanche Photodiodes for Charged Particles Timing}





\author[cern]{M.~Centis~Vignali\corref{corrauthor}}
\cortext[corrauthor]{Corresponding author}
\ead{matteo.centis.vignali@cern.ch}

\author[cern]{P.~Dias~De~Almeida}
\author[ubern]{L.~Franconi}
\author[cern,lip]{M.~Gallinaro}
\author[cern]{Y.~Gurimskaya}
\author[uprin]{B.~Harrop}
\author[ulan]{W.~Holmkvist}
\author[uprin]{C.~Lu}
\author[cern]{I.~Mateu}
\author[rmd]{M.~McClish}
\author[cern]{M.~Moll}
\author[upenn]{F.~M.~Newcomer}
\author[cern,usan]{S.~Otero~Ugobono}
\author[cern,uvirg]{S.~White}
\author[cern]{M.~Wiehe}

\address[cern]{CERN, Geneva, Switzerland}
\address[ubern]{University of Bern, Bern, Switzerland}
\address[lip]{LIP, Lisbon, Portugal}
\address[uprin]{Princeton University, Princeton, USA}
\address[ulan]{Lancaster University, Lancaster, UK}
\address[rmd]{Radiation Monitoring Devices, Watertown, USA}
\address[upenn]{University of Pennsylvania, Philadelphia, USA}
\address[usan]{Universidade de Santiago de Compostela, Santiago de Compostela, Spain}
\address[uvirg]{University of Virginia, Charlottesville, USA}

\begin{abstract}
The upgrades of ATLAS and CMS for the High Luminosity LHC (HL-LHC) highlighted physics objects timing as a tool to resolve primary interactions within a bunch crossing.
Since the expected pile-up is around 200, with an r.m.s. time spread of 180\,ps, a time resolution of about 30\,ps is needed.
The timing detectors will experience a 1-MeV neutron equivalent fluence of about $\Phi_{eq}=10^{14}$ and $10^{15}$\,cm$^{-2}$ for the barrel and end-cap regions, respectively.
In this contribution, deep diffused Avalanche Photo Diodes (APDs) produced by Radiation Monitoring Devices are examined as candidate timing detectors for HL-LHC applications.
To improve the detector's timing performance, the APDs are used to directly detect the traversing particles, without a radiator medium where light is produced.
Devices with an active area of $8\times8$\,mm$^2$ were characterized in beam tests.
The timing performance and signal properties were measured as a function of position on the detector using a beam telescope and a microchannel plate photomultiplier (MCP-PMT).
Devices with an active area of $2\times2$\,mm$^2$ were used to determine the effects of radiation damage and characterized using a ps pulsed laser. 
These detectors were irradiated with neutrons up to $\Phi_{eq}=10^{15}$\,cm$^{-2}$.
\end{abstract}


\end{frontmatter}

\section{Introduction}

The high luminosity upgrade of the CERN Large Hadron Collider (HL-LHC), foreseen to start in 2026, will provide an instantaneous luminosity of up to $7.5 \cdot 10^{34}$\,cm$^{-2}$s$^{-1}$ with a bunch spacing of 25\,ns, and an average pile-up of up to 200 collisions per bunch crossing\,\cite{hlLhcTecDesRep}.
To reduce the effects of pile-up on the physics analyses, both the ATLAS and CMS experiments are planning to implement dedicated systems to measure the time of arrival of minimum ionizing particles (MIPs) with an accuracy of about 30 ps\,\cite{cmsMIPtiming,atlasMIPtiming}.

By providing the time of arrival information of MIPs, these systems allow for the correct association of particles to their primary vertexes in the case the latter have a proximity in space that renders their separation impossible.
These timing detectors will be subjected to radiation levels corresponding to a 1-MeV neutrons fluence ($\Phi_{eq}$) of up to $4.9 \cdot 10^{15}$\,cm$^{-2}$ for the goal integrated luminosity of HL-LHC of 4000\,fb$^{-1}$.

This paper summarizes the characterization of deep diffused Avalanche Photo Diodes (APDs) produced by Radiation Monitoring Devices\,\cite{rmdAddress} used as timing detectors for charged particles.
Studies of these sensors as MIP timing detectors, using an AC-coupled readout, were performed previously and showed promising results\,\cite{white2014, vavra2017}.
The timing performance as well as the radiation hardness of these devices are addressed.

Section\,\ref{sec:ddApds} describes the devices used in this study.
Section\,\ref{sec:radHard} summarizes a radiation hardness study of devices with an active area of $2\times2$ mm$^2$.
Section\,\ref{sec:testbeam} contains the results of a beam test of devices with $8\times8$ mm$^2$ active area.
Finally, section\,\ref{sec:summary} summarizes the results presented in this paper.

\section{Deep Diffused APDs}
\label{sec:ddApds}

\begin{figure}
  \centering
  \includegraphics[width = 0.3 \columnwidth, trim=0.3cm 0.3cm 0.2cm 0.3cm, clip]{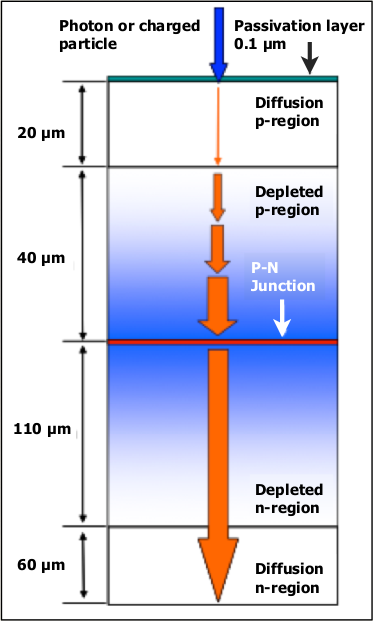}
  \caption{Schematic cross-sections of a deep diffused APD. The thickness of the depleted regions corresponds to a bias voltage of 1.8\,kV.}
  \label{fig:apdDia}
\end{figure}

Deep diffused APDs are silicon detectors based on charge multiplication.
They consist of a pn-junction operated in reverse bias.
Their doping profile and operation bias voltage allow for the electric field to reach values high enough to achieve impact ionization.
The typical bias voltage is close to 1800\,V, resulting in a gain of about 500 and a 150 $\mu$m thick depletion region.
The pn-junction lies a few tens of microns below the surface of the detector.
The fabrication process used to obtain such a configuration gives the name to the devices.
The depletion region does not reach the surfaces of the detector, therefore these APDs never reach full depletion.
A schematic cross section of the detector is shown in figure\,\ref{fig:apdDia}.
Further details about these devices can be found in\,\cite{mcclish2004, apdPatent, theoryDDAPD}.

Two kinds of APDs were used for the studies presented in this paper.
APDs with an active area of $2\times2$\,mm$^2$ and a die dimension of $3.1\times3.1$\,mm$^2$ were used for a radiation hardness study.
Devices with an active area of $8\times8$\,mm$^2$, on a $10\times10$\,mm$^2$ die, were characterized in a beam test.


\section{Radiation Hardness Study Using $2 \times 2$ mm$^2$ APDs}
\label{sec:radHard}

The radiation hardness of the APDs was determined using devices with an active area of $2 \times 2$ mm$^2$.
The APDs were irradiated with neutrons at the irradiation facility of the Jo\v{z}ef Stefan Institute in Ljubljana\,\cite{jsiIrrad}.
The accumulated fluences ranged from $\Phi_{eq} = 3 \cdot 10^{13}$ to $10^{15}$\,cm$^{-2}$.

The sensors were characterized using a pulsed infrared laser.
The details of the setup used and a broader range of results are given in\,\cite{centis2018, centis2019}.

\begin{figure}
  \centering
  \includegraphics[width = \columnwidth]{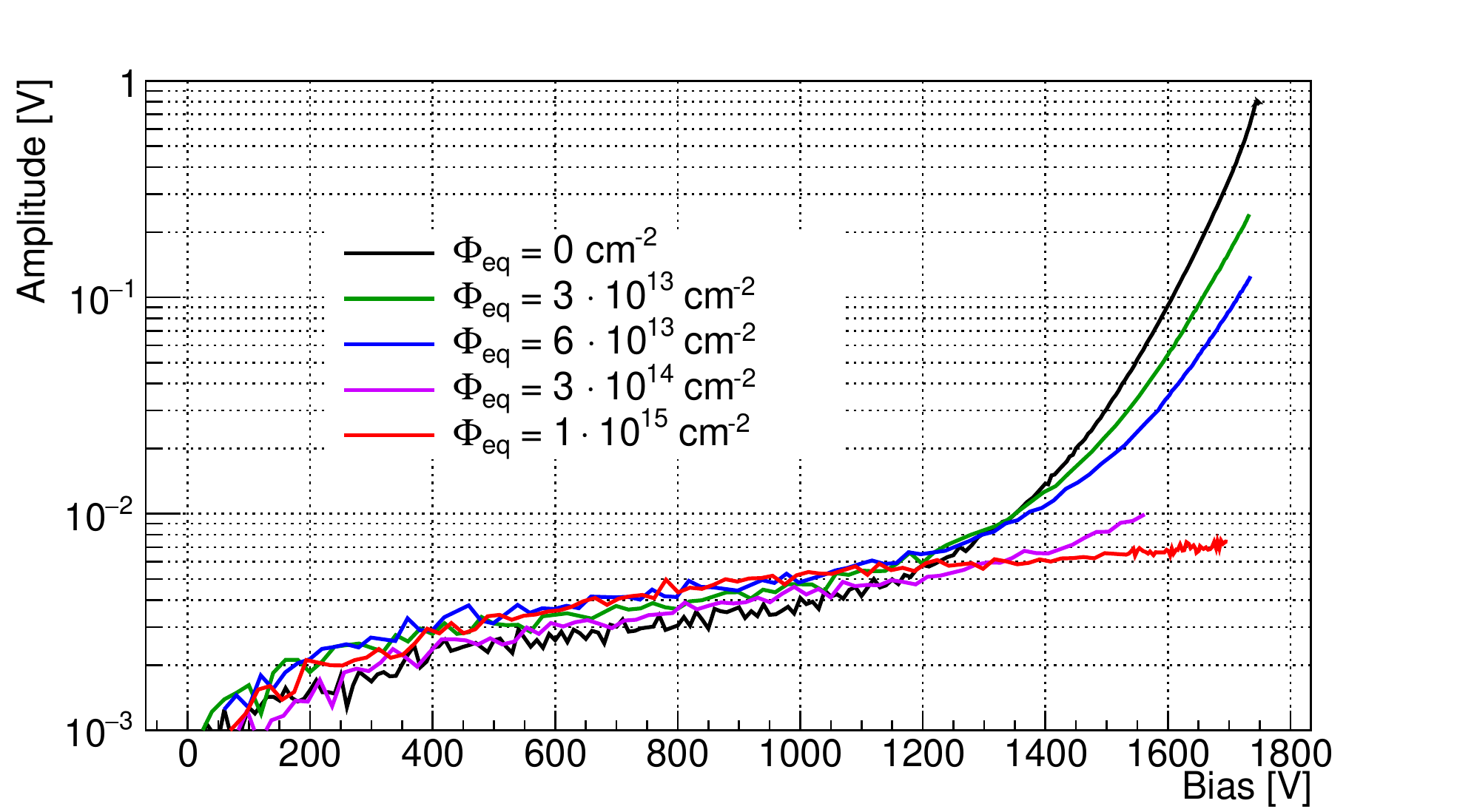}
  \caption{Amplitude of the APDs signal as a function of bias voltage and fluence. The measurements were performed at $-20^\circ$C. The laser intensity was 15\,MIP, and the amplification 10\,dB.}
  \label{fig:ampliIrrad}
\end{figure}

The evolution of the signal amplitude as a function of voltage and fluence can be seen in figure\,\ref{fig:ampliIrrad}.
The measurements were performed at $-20^\circ$C using laser pulses with an intensity corresponding to a charge deposit in the sensor of 15\,MIP.
The signal amplification was 10\,dB.
The irradiation decreases the gain of the detectors, for a given bias voltage.
The sensor irradiated to $\Phi_{eq} = 10^{15}$\,cm$^{-2}$ shows little to no gain.
The sensor irradiated to $\Phi_{eq} = 3 \cdot 10^{14}$\,cm$^{-2}$ is unstable above 1550\,V and was not measured at voltages above this value.
This fluence is considered the limit for the radiation hardness of these sensors since they cannot sustain the bias necessary for time of arrival measurements.

\begin{figure}
  \centering
  \includegraphics[width = \columnwidth]{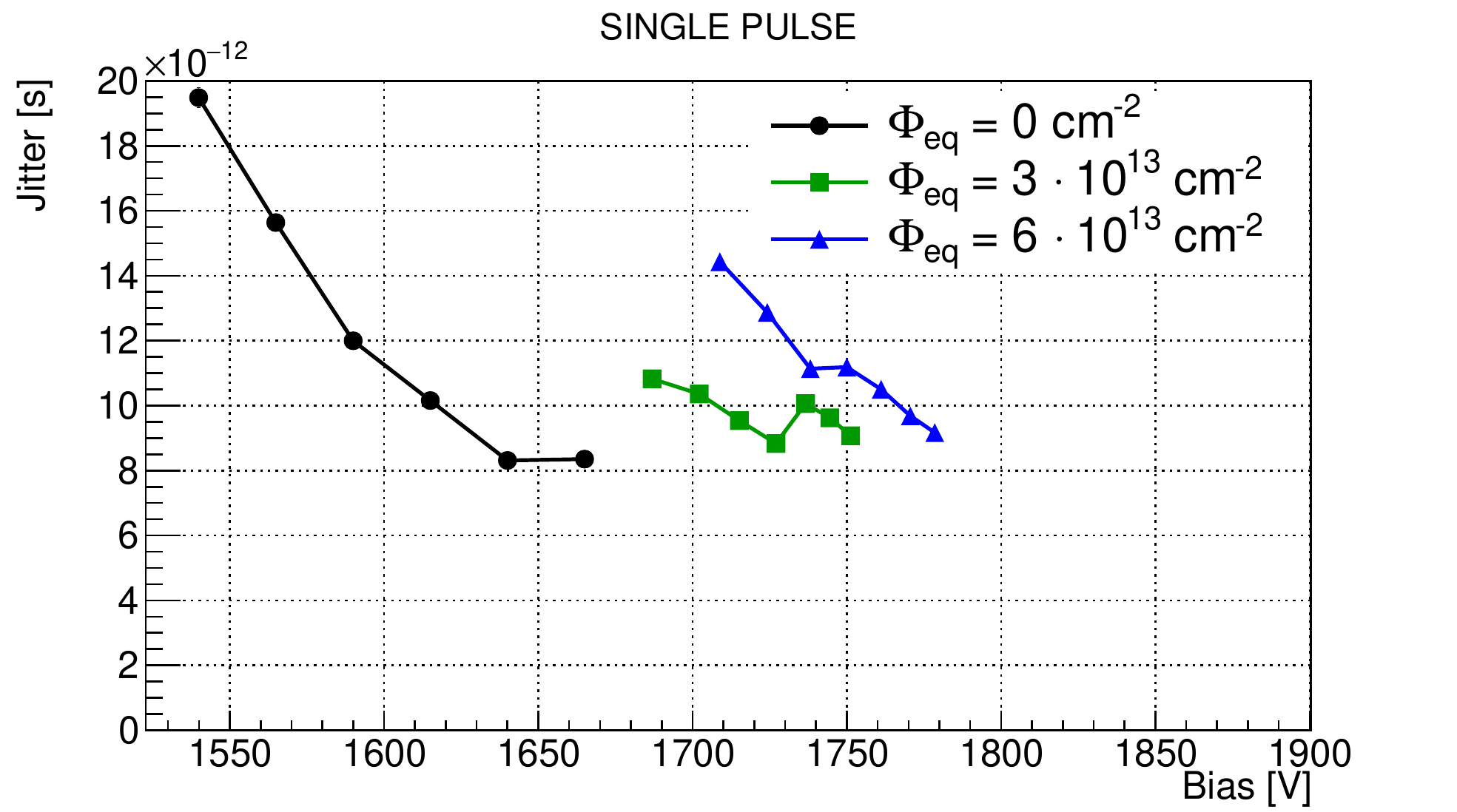}
  \caption{Jitter of the APDs signal as a function of bias voltage and fluence. The measurements were performed at $-20^\circ$C. The laser intensity was 0.8\,MIP, and the amplification 40\,dB.}
  \label{fig:timeResIrrad}
\end{figure}

The jitter of the APD signal was measured using a pulsed infrared laser with a 0.8\,MIP intensity and 40\,dB amplification.
The setup used for this measurement comprised an optical system that allows to project two light pulses on the sensor for each pulse generated by the laser.
The time difference between the pulses is 50\,ns, fixed by the optical system, and their amplitudes differ by less than 5\%.
The time resolution of the APDs can be extracted by measuring the time difference between the pulses.
In this configuration, no time reference detectors are needed, and the jitter of the laser does not influence the measurement.
Additional details about the setup and method used for this measurement can be found in\,\cite{centis2019}.
The jitter as a function of bias voltage and fluence is shown in figure\,\ref{fig:timeResIrrad}.
The jitter remains below 10\,ps for fluences up to $\Phi_{eq} = 6 \cdot 10^{13}$\,cm$^{-2}$, although the bias voltage necessary to achieve this performance increases with irradiation.

\begin{figure}[!t]
  \centering
  \includegraphics[width = \columnwidth]{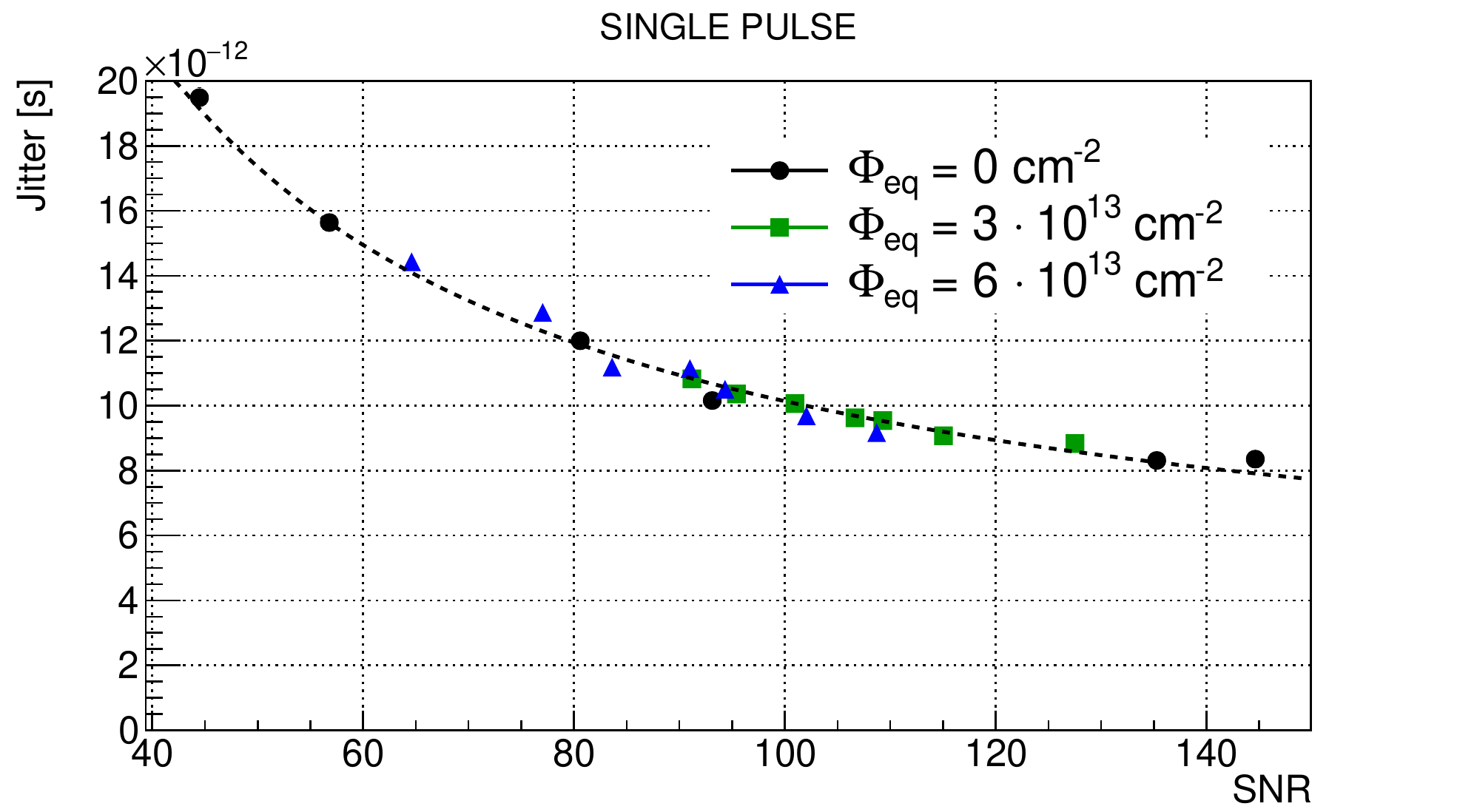}
  \caption{Jitter of the APDs signal as a function of signal to noise ratio. The measurements were performed under the same conditions of figure\,\ref{fig:timeResIrrad}. The dashed line represents a 1/SNR behavior fitted to the data.}
  \label{fig:timeResIrradSNR}
\end{figure}

The jitter was found to scale as one over the signal to noise ratio (1/SNR), as shown in figure\,\ref{fig:timeResIrradSNR}.

\section{Beam Test of $8 \times 8$ mm$^2$ APDs}
\label{sec:testbeam}

APDs with an active area of $8 \times 8$\,mm$^2$ were characterized in a beam test.
A metal layer was deposited on the detector surfaces to improve the uniformity of response.
The metal layer provides a DC-coupled readout of the detectors.
Additional details about the uniformity of response of these detectors can be found in\,\cite{centis2019}.
The main goals of the beam test were to measure the properties of the APD signal as a function of position on the detector, and the APD time resolution for MIPs.

The particles used were 100\,GeV muons at the H4 beam line at the CERN north area\,\cite{h4page}.
The beam test setup consisted of a microchannel plate photomultiplier (MCP-PMT), a beam telescope, trigger scintillators, and the APDs being tested.
The MCP-PMT was used as a time reference\,\cite{sohl2018}, while the beam telescope allowed for the reconstruction of the point where the particles traversed the APDs.
The trigger scintillators defined the acceptance of the setup.
The APDs were kept close to room temperature during the measurements.

The APDs were readout using a 2\,GHz, 40\,dB amplifier.
The signals of both the APDs and the MCP-PMT were digitized using a 2.5\,GHz, 10\,GSa/s oscilloscope, corresponding to a 100\,ps sampling interval.
The MCP-PMT signal was shaped using an attenuator and an amplifier, in order to acquire a few points on the signal's leading edge.
In this configuration the MCP-PMT time resolution is estimated to be about 10\,ps.

The results presented in this paper were obtained using one metallized APD operated at 1775\,V.

\begin{figure}
  \centering
  \includegraphics[width = \columnwidth]{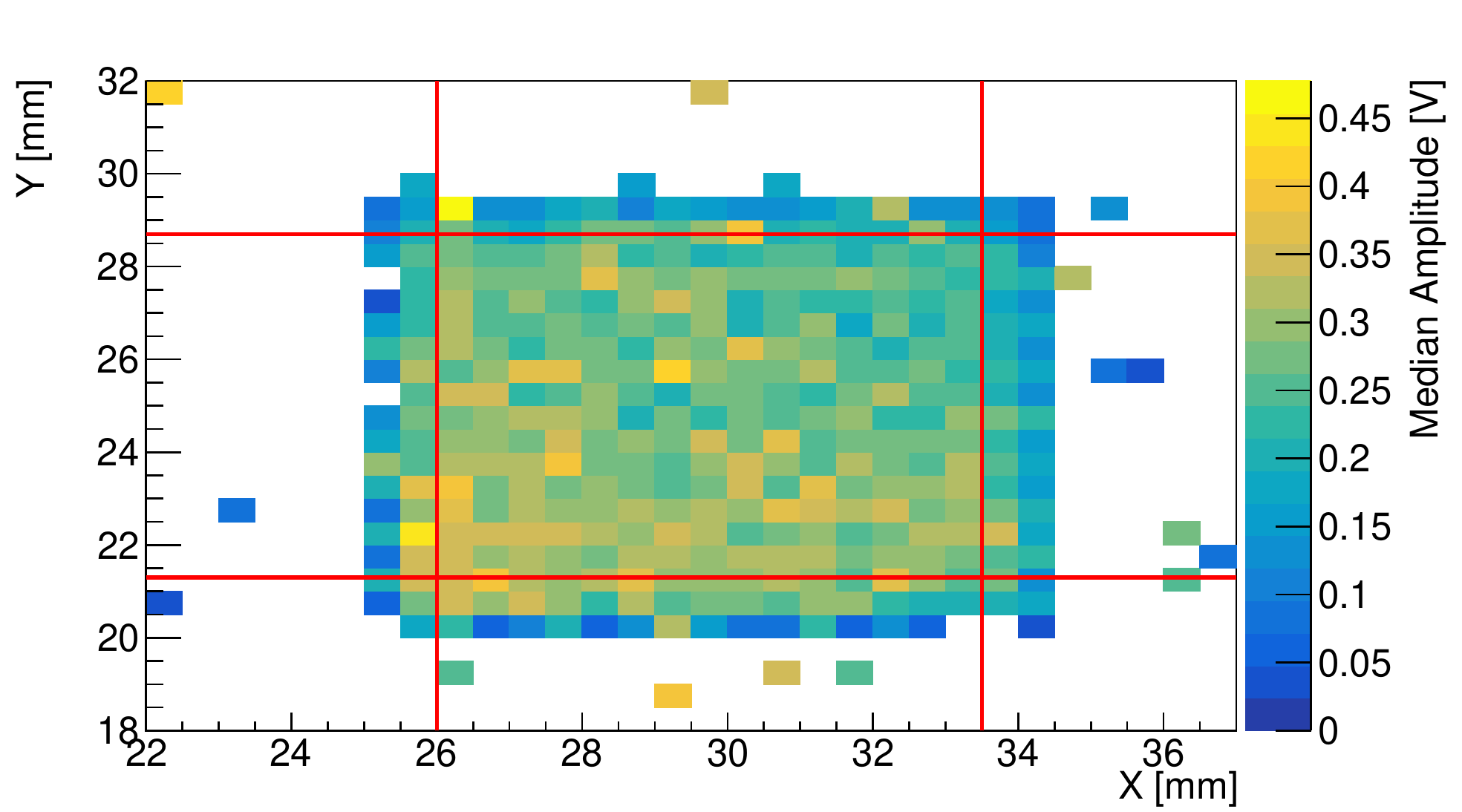}
  \caption{Median amplitude of the APD signal as a function of position. The APD bias was 1775\,V. The red lines represent the geometrical cuts used in the analysis.}
  \label{fig:ampliMap}
\end{figure}

The median amplitude of the APD signal as a function of position is shown in figure\,\ref{fig:ampliMap}.
The red lines represent the geometrical cuts used in the analysis.
The area considered for the analysis is $7.5\times7.4$\,mm$^2$.

\begin{figure}
  \centering
  \includegraphics[width = \columnwidth]{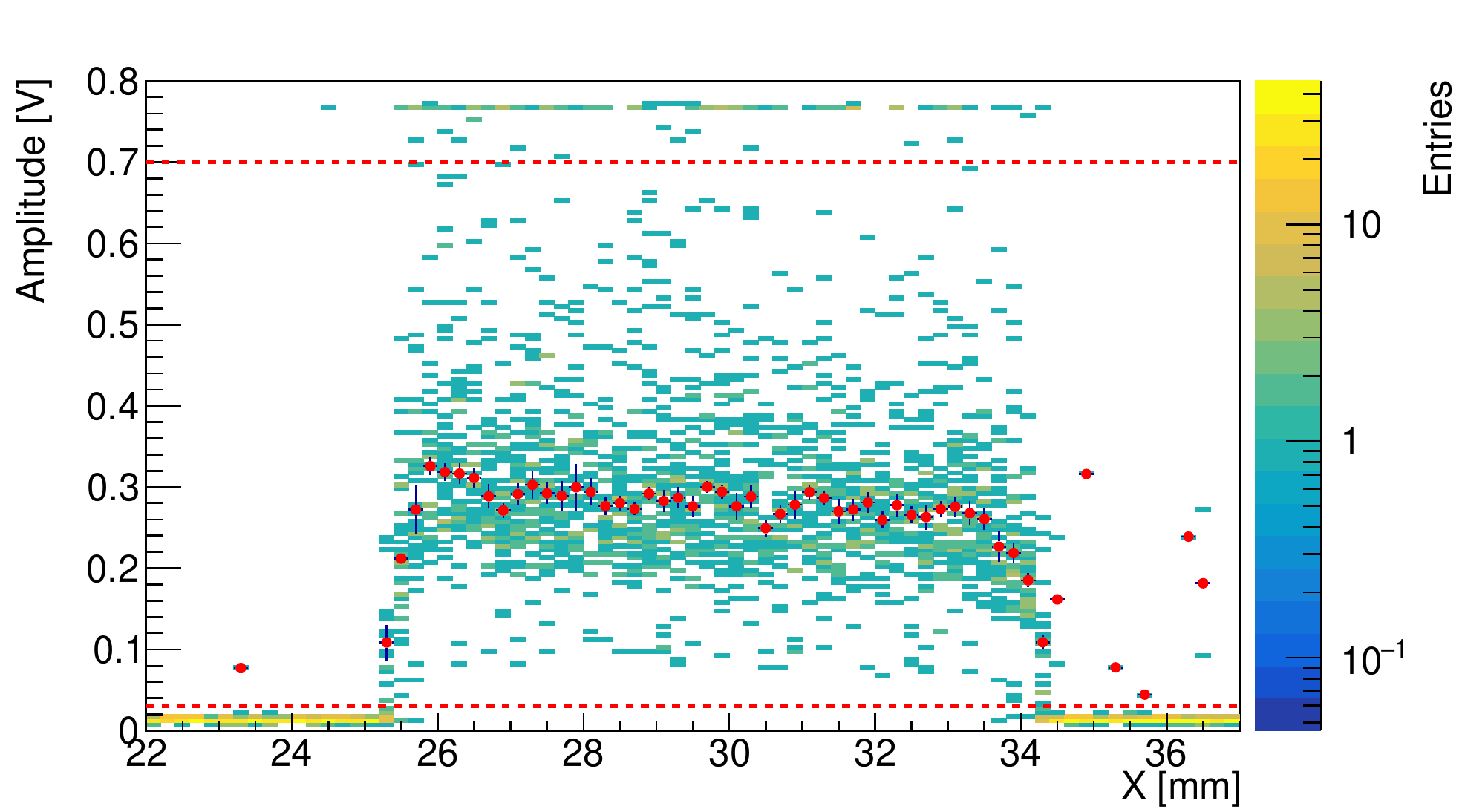}
  \caption{Amplitude of the APD signal as a function of position for the tracks contained between the horizontal lines of figure\,\ref{fig:ampliMap}. The red points represent the median amplitude, while the dashed lines represent the cuts used to calculate the median. The APD bias was 1775\,V.}
  \label{fig:ampliSlice}
\end{figure}

Figure\,\ref{fig:ampliSlice} shows the amplitude of the APD signal as a function of position for the tracks contained between the horizontal lines in figure\,\ref{fig:ampliMap}.
The red dots represent the median amplitude as a function of position.
The dashed lines represent the cuts in amplitude used to calculate the median: a threshold of 30\,mV is applied, and the amplitudes considered are limited to 0.7\,V to exclude the events saturating the oscilloscope scale.
The amplitude is uniform over the detector, for the area considered in the analysis.

Using a threshold of 30\,mV and the information from the telescope, the detection efficiency can be measured.
The efficiency is in this case defined as the number of events with a signal amplitude above threshold divided by the number of tracks reconstructed using the telescope.
Within the geometrical cuts, the detection efficiency is above 99\%.

\begin{figure}
  \centering
  \includegraphics[width = \columnwidth]{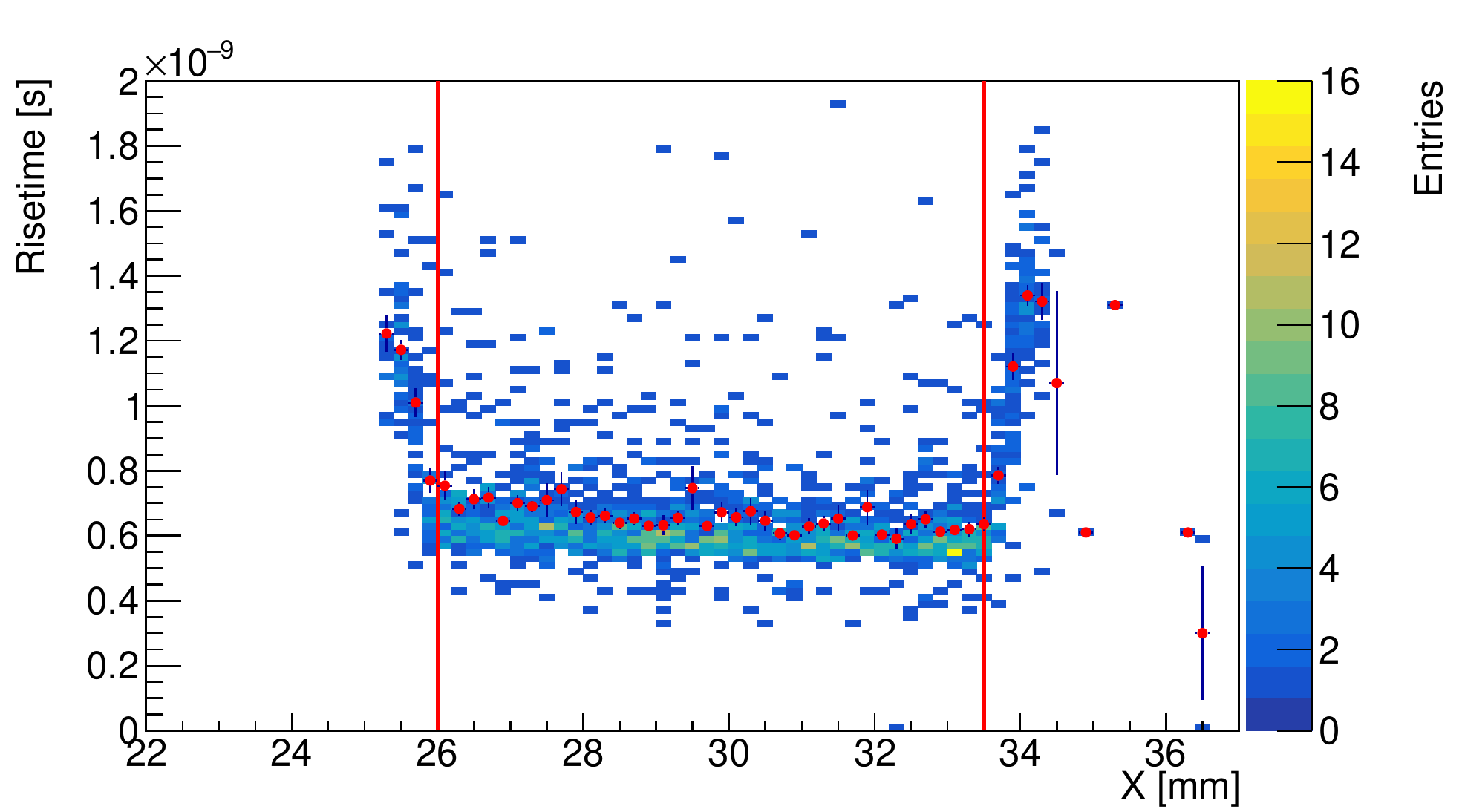}
  \caption{20 to 80\% rise time of the APD signal as a function of position for the tracks contained between the horizontal lines of figure\,\ref{fig:ampliMap}. The red points represent the average rise time as a function of position, while the vertical lines represent the geometrical cuts. The APD bias was 1775\,V.}
  \label{fig:riseTimeSlice}
\end{figure}

The 20 to 80\% rise time of the APD signal as a function of position is shown in figure\,\ref{fig:riseTimeSlice}.
The selected tracks are contained within the horizontal lines of figure\,\ref{fig:ampliMap}.
Only events fulfilling the amplitude cuts shown in figure\,\ref{fig:ampliSlice} are used for rise time measurements.
The red points represent the average rise time as a function of position.
The vertical lines represent the geometrical cuts.
The rise time is uniform over the area selected for the analysis, while the edges of the detector show higher rise time values.

\begin{figure}
  \centering
  \includegraphics[width = \columnwidth]{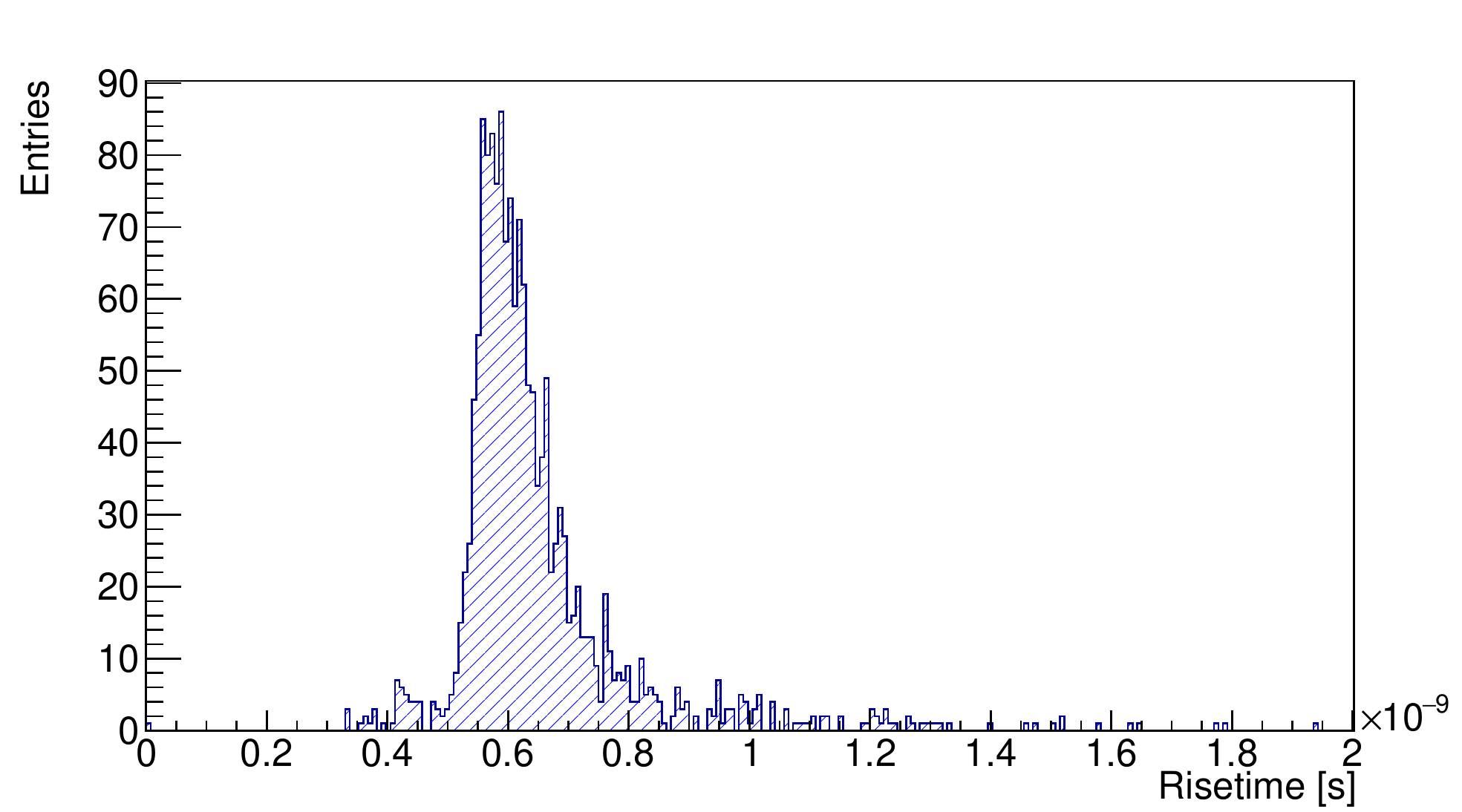}
  \caption{20 to 80\% rise time of the APD signal for the tracks fulfilling the geometrical and amplitude cuts. The APD was biased to 1775\,V.}
  \label{fig:riseTimeDistr}
\end{figure}

The distribution of the 20 to 80\% rise time for the tracks fulfilling the geometrical (fig.\,\ref{fig:ampliMap}) and amplitude (fig.\,\ref{fig:ampliSlice}) cuts is shown in figure\,\ref{fig:riseTimeDistr}.
The most probable value of the distribution is around 600\,ps.
The tail toward higher values is not yet understood.

\begin{figure}
  \centering
  \includegraphics[width = \columnwidth]{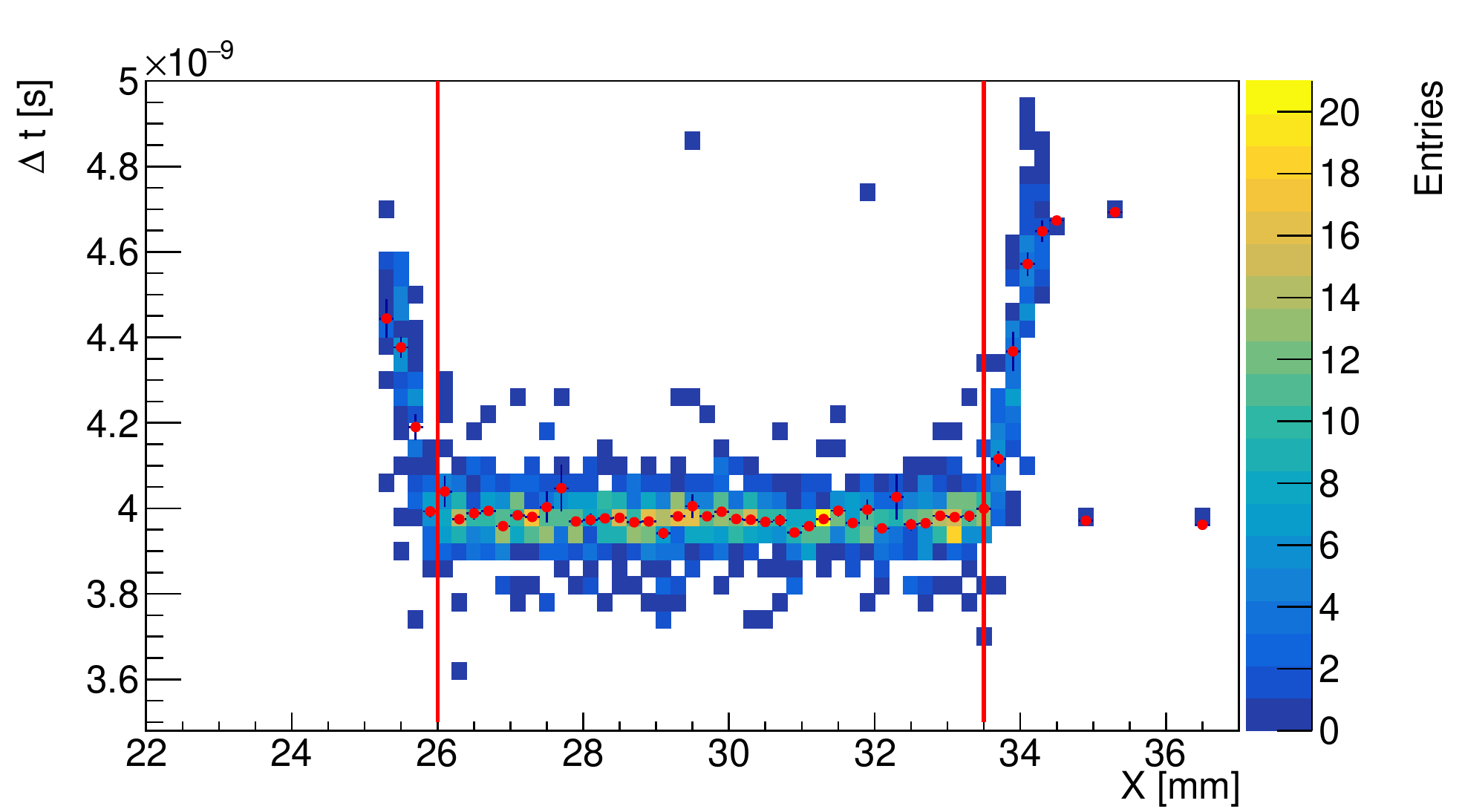}
  \caption{Time of arrival as a function of position for the tracks contained between the horizontal lines of figure\,\ref{fig:ampliMap}. The red points represent the average $\Delta t$ as a function of position, while the vertical lines represent the geometrical cuts. The APD bias was 1775\,V.}
  \label{fig:toaSlice}
\end{figure}

The time of arrival $\Delta t$ is defined as the difference in the time between the MCP-PMT and APD signals.
A constant fraction discriminator (CFD) algorithm was used, interpolating between two points of the digitized waveform to determine the threshold crossing.
The CFD threshold applied to the APD was 0.3, while the one on the MCP-PMT was 0.5.
The values of $\Delta t$ as a function of position are shown in figure\,\ref{fig:toaSlice} for the tracks contained within the horizontal lines of figure\,\ref{fig:ampliMap}.
Only events fulfilling the amplitude cuts shown in figure\,\ref{fig:ampliSlice} are used for $\Delta t$ measurements.
The red points represent the average $\Delta t$ as a function of position, and the vertical lines the geometrical cuts.
The time of arrival is uniform over the detector, for the area considered in the analysis.
The edges of the detector show higher $\Delta t$ values; this behavior is similar to the one seen for the rise time.

\begin{figure}
  \centering
  \includegraphics[width = \columnwidth]{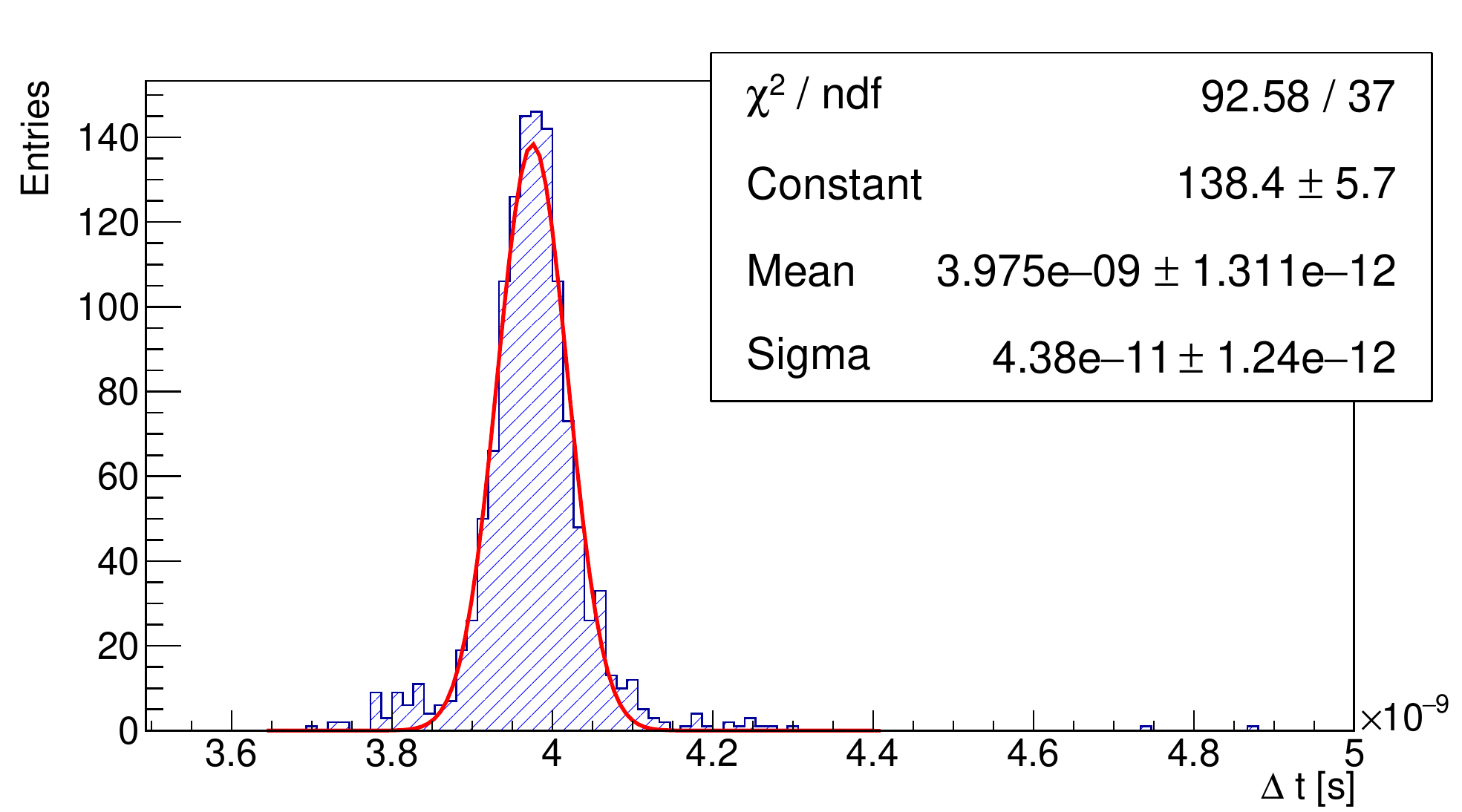}
  \caption{Time of arrival distribution for the tracks fulfilling the geometrical and amplitude cuts. The APD bias was 1775\,V.}
  \label{fig:toaDistr}
\end{figure}

The time of arrival distribution for the tracks fulfilling the geometrical (fig.\,\ref{fig:ampliMap}) and amplitude (fig.\,\ref{fig:ampliSlice}) cuts is shown in figure\,\ref{fig:toaDistr}.
By describing the distribution with a Gauss function, a time resolution of $44 \pm 1$\,ps is obtained for the MCP-PMT APD system over the $7.5\times7.4$\,mm$^2$ area considered for this study.
The time resolution degradation effect due to the MCP-PMT is expected to be negligible.

It is worth to remark that the readout scheme used for the APDs in the beam test was not optimal for timing measurements, and that the performance of the detectors is expected to improve with an improved readout scheme.
During the beam test, the readout was done connecting the detector in series with the amplifier.
The high voltage is dropped on the detector, with the amplifier directly connected to the detector side close to ground potential.
In this configuration, the RC constant that contributes to the leading edge of the pulse is affected by: the detector capacitance, the input impedance of the amplifier, and the impedance of the cable that connects the detector to the high voltage power supply.
The readout can be improved by using a bias tee, connecting the detector in parallel to the amplifier through a capacitor.
In this case, the impedance of the cable between detector and high voltage supply does not affect the RC constant, resulting in a steeper leading edge of the signal and therefore improving the time resolution.
The mechanical constrains of the setup unfortunately did not allow to use a bias tee for the APD readout.

\section{Summary}
\label{sec:summary}

Deep diffused APDs were proposed as MIP timing detectors.
The radiation hardness of these devices was assessed using devices with a $2\times2$\,mm$^2$ active area.
The jitter of these detectors was measured using a pulsed infrared laser and it remains below 10\,ps up to a fluence $\Phi_{eq} = 6 \cdot 10^{13}$\,cm$^{-2}$.
The detectors are expected to operate up to a fluence $\Phi_{eq} \approx 10^{14}$\,cm$^{-2}$.
Devices with an active area of $8\times8$\,mm$^2$ and a DC-coupled readout were characterized in a beam test.
The time resolution of these detectors was found to be $44\pm1$\,ps over a $7.5\times7.4$\,mm$^2$ area, at a bias of 1775\,V.
This performance is expected to be improved by changing the readout scheme.

\section*{Acknowledgments}

The work summarized in this paper has been performed within the framework of the RD50 collaboration.
This project has received funding from the European Union’s Horizon 2020 Research and Innovation programme under Grant Agreement no.\ 654168.
The authors wish to thank J.~Bronuzzi for the help during the clean room operations and in the development of the recipe for the metallization of the devices.
The authors would like to thank the RD51 and PICOSEC collaborations for the possibility to participate in the May and August 2018 beam tests, and especially E.~Oliveri, F.~Iguaz Gutierrez, and L.~Sohl.
The authors wish to thank F.~Garcia Fuentes and R.~De~Oliveira for the help in preparing the sensors and PCBs.

\bibliography{bibliography}

\end{document}